\documentclass[twocolumn,english,reprint,aps,prl]{revtex4-1}
\usepackage[T1]{fontenc}
\usepackage[latin9]{inputenc}
\usepackage{color}
\usepackage{babel}
\usepackage{amsmath}
\usepackage{graphicx}
\usepackage{wasysym}
\usepackage[unicode=true,
 bookmarks=false,
 breaklinks=false,pdfborder={0 0 1},backref=false,colorlinks=true]
 {hyperref}
\hypersetup{
 linkcolor=blue,citecolor=blue,urlcolor=blue,linkcolor=blue,citecolor=blue,urlcolor=blue}
\usepackage{breakurl}

\makeatletter

\usepackage{babel}

\usepackage{babel}

\usepackage{babel}
\usepackage{babel}

\usepackage{calrsfs}
\DeclareMathAlphabet{\pazocal}{OMS}{zplm}{m}{n}

\usepackage{wasysym}
\usepackage{babel}
\usepackage{amsfonts}

\usepackage{graphics}

\usepackage{babel}

\makeatother

\begin{document}

\title{Enhancement of superconductivity with external phonon squeezing}

\author{Andrey Grankin}

\author{Mohammad Hafezi}

\author{Victor M. Galitski}

\affiliation{Joint Quantum Institute, University of Maryland, College Park, MD
20742, USA}

\affiliation{Condensed Matter Theory Center, Department of Physics, University
of Maryland, College Park, MD 20742, USA}
\begin{abstract}
Squeezing of phonons due to the non-linear coupling to electrons is
a way to enhance superconductivity as theoretically studied in a recent
work {[}Kennes \textit{et al.} Nature Physics \textbf{13}, 479 (2017){]}.
We study quadratic electron-phonon interaction in the presence of
phonon pumping and an additional external squeezing. Interference
between these two driving sources induces a phase-sensitive enhancement
of electron-electron attraction, which we find as a generic mechanism
to enhance any boson-mediated interactions. The strongest enhancement
of superconductivity is shown to be on the boundary with the dynamical
lattice instabilities caused by driving. We propose several experimental
platforms to realize our scheme. 
\end{abstract}
\maketitle
Optical excitation of infrared-active (IR) phonon modes allows for
ultrafast pumping of solid-state systems into non-equilibrium states
exhibiting a wide range of exotic properties \citep{BAH17}. In particular,
this enables manipulation of magnetic states \citep{FTW11}, charge
orders \citep{RHW11}, and superconductivity \citep{MCN16,MFL15}.
A possible explanation of the transient enhancement observed in \citep{MCN16}
is the parametric driving of Raman phonons by IR phonons \citep{KBR16,BKM17}.
Enhanced \citep{QML18} quantum fluctuations of phonon modes lead
to a stronger phonon-mediated attraction between electrons thus enhancing
the superconductivity.

\begin{figure}[!h]
\centering \includegraphics[scale=0.45]{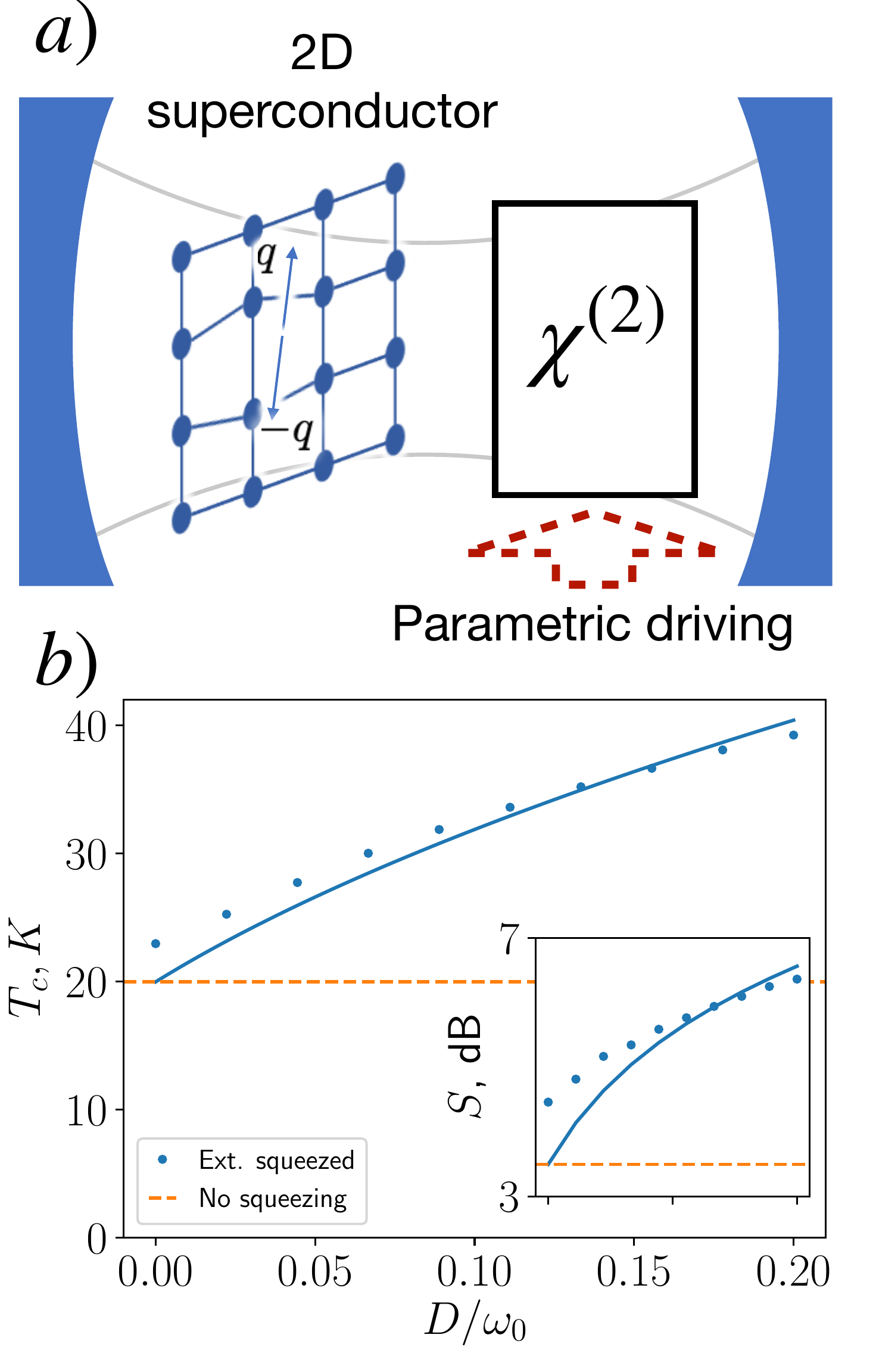} \caption{Steady state enhancement of superconducting transition temperature
$T_{c}$. (a) Sketch of the setup: 2D superconductor on an optomechanical
membrane interacting with the squeezed light produced by the optical
parametric oscillator (OPO). (b) Superconducting transition temperature
as function of the parametric driving strength $D$ for optimal value
of the detuning $\delta$, assuming the phonon driving is fixed at
$\alpha=0.2\omega_{0}$: parametrically driven phonons (blue), driven
phonons without external squeezing (orange dashed). The corresponding
phonon excess noise \textbf{$S\equiv N^{-1}\sum_{{\bf {q}}}\langle\hat{\phi}_{{\bf {q}}}\hat{\phi}_{-{\bf {q}}}\rangle$}
is shown on the inset.}
\label{Fig1} 
\end{figure}

Another proposed mechanism of transient superconductivity enhancement
is based on a nonlinear coupling between electrons and phonons \citep{KWR17,S17}.
In this case, the squeezing of phonons is generated directly by the
electron-phonon interaction, since the coupling is quadratic in the
phonon operator. As shown in \citep{S17} due to this nonlinearity,
the effective interaction is enhanced proportionally to the coherent
excitation rate of the phonon mode.

In this Letter, we clarify the role of squeezing and study the possibility
of enhancement of superconductivity by an additional external parametric
drive. We consider a model that combines both ingredients -- the
linear and parametric driving of phonons that are non-linearly coupled
to electrons. In order to illustrate the influence of parametric driving
on a bosonic degree of freedom $[a,a^{\dagger}]=1$ and introduce
a related terminology, we define the Hamiltonian of a parametrically
driven harmonic oscillator \cite{WM07} with a bare frequency $\omega_{0}$
as $H_{\text{PO}}=\omega_{0}\hat{a}^{\dagger}\hat{a}-D(\hat{a}^{2}e^{2i\omega_{p}t}+\hat{a}^{\dagger2}e^{-2i\omega_{p}t})/2$
where $D$ stands for the squeezing strength, $\omega_{p}$ is the
parametric driving frequency. One can show that the expectation value
of the quadrature $\hat{X}_{\theta}\equiv\hat{a}e^{i\theta+i\omega_{p}t}+\text{H.c}.$
decreases for $\theta=\pi$ and increases for $\theta=0$, with respect
to a local oscillator. We refer to these quadratures as to ``squeezed''
and ``anti-squeezed'', respectively. Similarly, the corresponding
retarded correlation function of these quadratures, which determines
the strength of mediated interaction by these bosonic modes, can decrease
or increase in a phase-sensitive fashion (see the Supplementary Material
for spins as an illustrative example). Consequently, the squeezing
of proper phonon quadrature can significantly amplify the photon-mediated
electron-electron interaction. Moreover, the parametric drive can
soften the photon modes that further amplifies the electronic interaction.
In this work, we demonstrate that such these amplifications lead to
the enhancement of superconductivity, by analytically and numerically
employing the Migdal-Eliashberg theory. The superconducting critical
temperature $T_{c}$ is shown in Fig.~\ref{Fig1}~(b) as function
of external phonon squeezing rate $D$. The external parametric drive
that is the crucial element of our proposal can be achieved by either
exploiting intrinsic photon-phonon coupling nonlinearities \citep{CNF18}
or using an parametric optical amplifier in an optical cavity to produce
squeezed light \citep{GH16} as schematically shown in Fig.~\ref{Fig1}
(a).

To be specific, we study a 2-dimensional superconductor interacting
with an infrared-active optical phonon mode and consider the coupling
to be quadratic in phonon operator \citep{MCN16,KWR17,S17}. Linear
coupling terms can also be present without affecting the results below.
The full Hamiltonian of the system reads as $\hat{H}_{\text{full}}=\hat{H}_{\text{p}}+\hat{H}_{\text{e}}+\hat{H}_{\text{e-p}}+\hat{V}(t)$:
\begin{align}
\hat{H}_{\text{p}} & =\underbrace{\sum_{{\bf q}}\omega_{{\bf q}}\hat{a}_{{\bf q}}^{\dagger}\hat{a}_{{\bf q}}}_{H_{\text{p}}}+\underbrace{\sum_{{\bf {k},\sigma}}(\epsilon_{{\bf {k}}}-\mu)\hat{c}_{{\bf k,\sigma}}^{\dagger}\hat{c}_{{\bf k,\sigma}}}_{H_{\text{e}}}\nonumber \\
 & +\underbrace{\frac{g}{N}\sum_{\sigma,{\bf k},{\bf q},{\bf q}^{\prime}}c_{{\bf {k}+q-q^{\prime},\sigma}}^{\dagger}\hat{c}_{{\bf {k},\sigma}}\hat{\phi}_{{\bf {q}}}\hat{\phi}_{-{\bf {q}^{\prime}}}}_{H_{\text{e-p}}},\label{eq:H1}
\end{align}
where we $N$ denotes the total number of lattice sites, ${\bf q}$
is the lattice quasi-momentum vector, $\omega_{{\bf {q}}}$ and $\epsilon_{{\bf {k}}}$
respectively stand for the phonon and electron dispersions, $\hat{c}_{{\bf {k},\sigma}}$
is the electron annihilation operator and $\hat{\phi}_{{\bf {q}}}\equiv\hat{a}_{{\bf {q}}}+\hat{a}_{-{\bf {q}}}^{\dagger}$
is the phonon displacement field operator. Phonons are linearly and
parametrically driven at frequency $\omega_{p}$ with the corresponding
driving strengths $\alpha$ and $D_{{\bf {q}}}$. The external driving
Hamiltonian reads:

\begin{equation}
\hat{V}\left(t\right)=2\alpha\cos(\omega_{p}t+\theta_{\alpha})\hat{\phi}_{0}+\sum_{{\bf q}}D_{{\bf q}}\cos(2\omega_{p}t+\theta_{D})\hat{\phi}_{{\bf q}}\hat{\phi}_{-{\bf q}}
\end{equation}
where $\theta_{\alpha,D}$ are the relative phases of linear and parametric
drivings. As we discuss below, these phases allow
one to control the strength of coupling to electrons. We note that
the model becomes dynamically unstable at strong parametric drive
$D_{{\bf {q}}}$ \citep{WM07}. This is manifested in the exponential
growth of the phonon displacement $\langle\hat{\phi}_{{\bf {q}}}\rangle$,
as function of time. Therefore, we impose $\omega_{{\bf {q}}}-\omega_{p}\geq D_{{\bf {q}}}$
to avoid such an instability.

\begin{figure}
\begin{centering}
\includegraphics[scale=0.35]{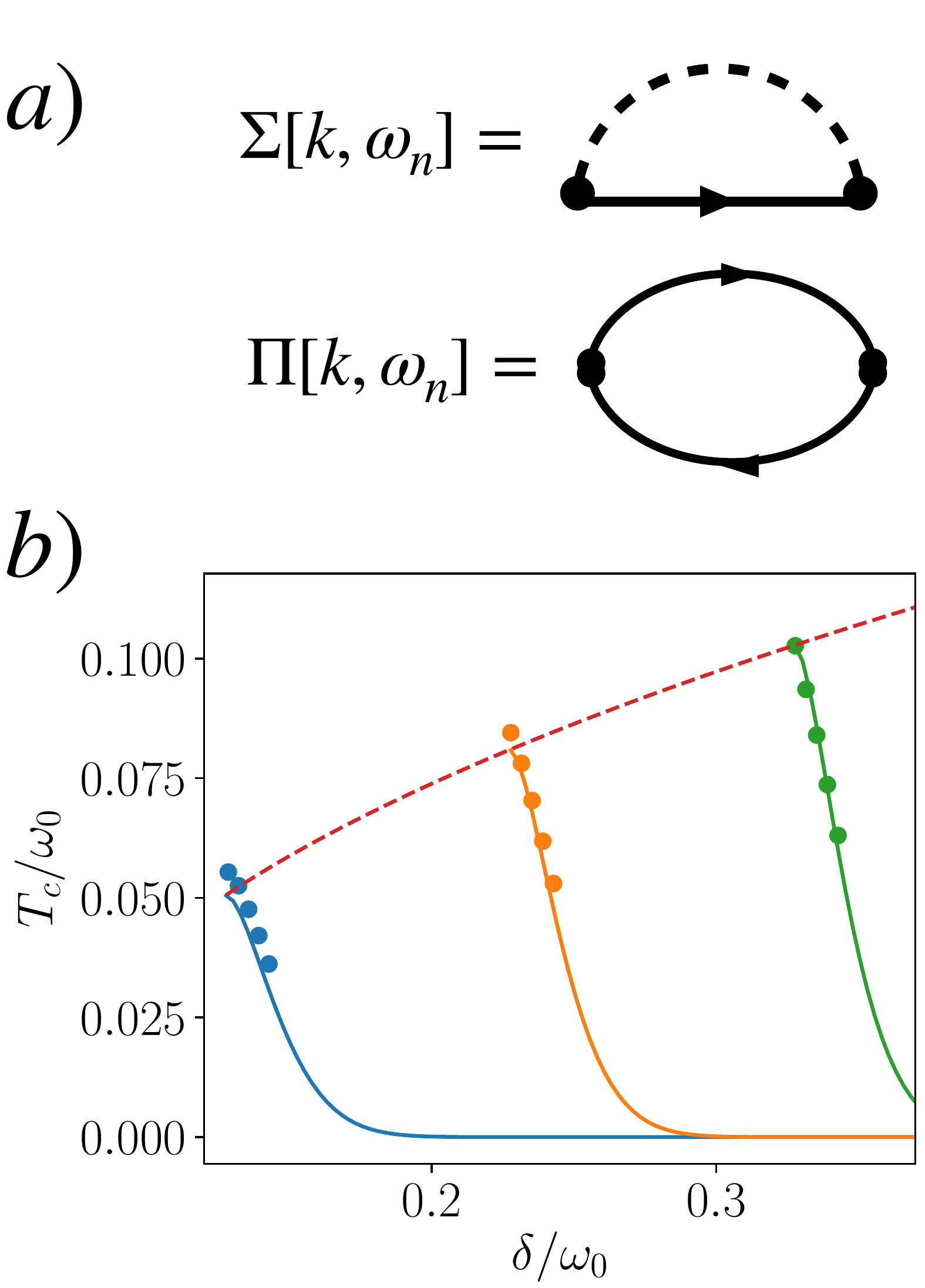} 
\par\end{centering}
\caption{Enhancement of superconductivity. a) Electronic and phononic self-energies
within Migdal-Eliashberg theory. Solid and dashed lines respectively
stand for the fully renormalized electron and phonon propagators.
b) Superconducting $T_{c}$ as function of the detuning $\delta$:
configuration without squeezing $D=0$ (blue), $D=0.1\omega_{0}$
(orange) and $D=0.2\omega_{0}$ (green) and optimal values (red dashed).
Dotted curves correspond to the exact numerical solution of the Migdal-Eliashberg
equations (\ref{eq:G}-\ref{eq:Phi}) on a discretized momentum-frequency
lattice. }

\label{Fig2} 
\end{figure}

The external drive induces a finite expectation value of the zero-momentum
phonon mode $\langle\hat{\phi}_{0}\rangle$. We treat this in terms
of mean-field theory and keep quadratic fluctuations. We perform two
unitary transformations of the Hamiltonian Eq.~(\ref{eq:H1}). First,
we consider the frame, rotating at the phonon driving frequency $\omega_{p}$,
which transforms bosonic variables as $\hat{a}_{{\bf {q}}}\rightarrow\hat{a}_{{\bf {q}}}e^{-i\omega_{p}t}$.
Second, we perform a shift of the zero-momentum bosonic variables
$\hat{a}_{{\bf {q}}}\rightarrow\hat{a}_{{\bf {q}}}+\bar{a}_{0}\delta_{{\bf {q},0}}$,
where $\bar{a}_{0}$ denotes the adiabatic steady state coherence
to the lowest order in $1/\omega_{p}$ (see SM): 
\begin{equation}
\bar{a}_{0}\equiv\alpha\frac{D_{0}e^{i\left(\theta_{\alpha}-\theta_{D}\right)}-\delta_{0}e^{-i\theta_{\alpha}}}{\delta_{0}^{2}-D_{0}^{2}}\label{a0}
\end{equation}
Finally, we perform the rotating-wave approximation (RWA) and discard
the rotating at frequencies $\propto2\omega_{p}$,  by assuming that
the driving frequency $\omega_{p}$ is the largest energy scale in
the system. As shown in the supplementary material, the effective
coupling in the model is maximized for the following choice of driving
phases $\theta_{D}=\pi,\theta_{\alpha}=0$. This choice corresponds
to ``anti-squeezing'' of the quadrature to which electrons are coupled.
With these approximations and neglecting all non-linear and rotating
contributions, the phonon Hamiltonian and the electron-phonon Hamiltonians
are transformed as:

\begin{align}
\hat{H}_{\text{ph}} & =\sum_{{\bf q}}\delta_{{\bf q}}\hat{a}_{{\bf q}}^{\dagger}\hat{a}_{{\bf q}}-\sum_{{\bf {q}}}\frac{D_{{\bf {q}}}}{2}(\hat{a}_{{\bf {q}}}\hat{a}_{-{\bf {q}}}+\hat{a}_{{\bf q}}^{\dagger}\hat{a}_{-{\bf q}}^{\dagger})\label{eq:H_ph}\\
\hat{H}_{\text{int}} & =\frac{g_{\text{eff}}}{\sqrt{N}}\sum_{\sigma,q}\hat{c}_{{\bf k+q,\sigma}}^{\dagger}\hat{c}_{{\bf k,\sigma}}(\hat{a}_{{\bf q}}+\hat{a}_{-{\bf {q}}}^{\dagger}),\label{eq:H_int}
\end{align}
where the detuning $\delta_{{\bf {q}}}\equiv\omega_{{\bf {q}}}-\omega_{p}$
and the effective electron-phonon coupling is $g_{\text{eff}}=2g|\bar{a}_{0}|/\sqrt{N}$.
Eq.~\eqref{eq:H_ph} is equivalent to a non-degenerate multimode
parametric oscillator \citep{WM07} below the parametric instability
threshold for $\delta_{{\bf q}}\geq D_{{\bf q}}$, which can be diagonalized
by means of the Bogolyubov transformation $\hat{a}_{{\bf q}}=\cosh(r_{{\bf {q}}})\hat{b}_{{\bf {q}}}+\sinh(r_{{\bf {q}}})\hat{b}_{-{\bf {q}}}^{\dagger}$
with $\zeta_{{\bf {q}}}=2^{-1}\text{arctanh}(D_{{\bf {q}}}/\delta_{{\bf {q}}})$.
We find $\hat{H}_{\text{ph}}=\sum_{{\bf {q}}}\sqrt{\delta_{{\bf {q}}}^{2}-D_{{\bf {q}}}^{2}}\hat{b}_{{\bf {q}}}^{\dagger}\hat{b}_{{\bf {q}}}$
and $\hat{H}_{\text{int}}=N^{-1/2}g_{\text{eff}}\sum_{\sigma,{\bf q}}e^{\zeta_{{\bf {q}}}}\hat{c}_{{\bf {k}+q,\sigma}}^{\dagger}\hat{c}_{{\bf {k},\sigma}}(\hat{b}_{{\bf {q}}}+\hat{b}_{-{\bf {q}}}^{\dagger})e^{i{\bf q}{\bf r}_{i}}$.
Close to the parametric instability $D_{{\bf {q}}}\sim\delta_{{\bf {q}}}$,
the coupling scales as $e^{\zeta_{{\bf {q}}}}\propto\left(1-D_{{\bf {q}}}/\delta_{{\bf {q}}}\right)^{-1/4}$.
Eqs.~(\ref{eq:H_ph}, \ref{eq:H_int}) are therefore equivalent to
a conventional Holstein model \citep{AGD63} with the softened phonons
and an enhanced electron-phonon coupling. As we show below, the combination
of these factors can lead to an enhanced $T_{c}$ compared to the
configuration without squeezing.

In order to show this enhancement, we  consider the squeezed electron-phonon
model of Eqs. (\ref{eq:H_ph}, \ref{eq:H_int}) within the equilibrium
Migdal-Eliashberg (ME) theory \citep{M20,ENH18} and provide an estimate
of the superconducting phase transition temperature $T_{c}$. ME theory
relies on the Migdal theorem  that allows one to neglect vertex corrections
to the electron Green's function provided they are much faster than
phonons. In case of the effective Holstein model (\ref{eq:H_ph},
\ref{eq:H_int}), this is characterized by $\sqrt{\delta_{{\bf {q}}}^{2}-D_{{\bf {q}}}^{2}}\ll E_{\text{F}}$,
where $E_{\text{F}}$ is the Fermi energy. The remaining equations
for the electronic and phonon self-energies form a closed set of equations,
which can be solved self-consistently, and we consider the formulation
of the theory above the critical temperature $T\apprge T_{c}$.

We start by defining the fully-renormalized imaginary-time propagators
${\cal G}_{{\bf {k}}}^{-1}=i\omega_{n}-(\epsilon_{{\bf {k}}}-\mu)-\Sigma_{{\bf {k}}}(i\omega_{n})$,
${\cal D}_{{\bf {k}}}^{-1}={\cal D}_{{\bf {k}}}^{\left(0\right)-1}(i\omega_{m})-\Pi_{{\bf {k}}}(i\omega_{n})$:
where the unperturbed squeezed phonon propagator is ${\cal D}_{{\bf {q}}}^{\left(0\right)}(i\omega_{m})=-2(\delta_{{\bf {q}}}+D_{{\bf {q}}})/(\omega_{m}^{2}+\delta_{{\bf {q}}}^{2}-D_{{\bf {q}}}^{2})$
and $\omega_{n}=\pi(2n+1)/\beta$, $\omega_{m}=2\pi m/\beta$, $m,n\in\mathbb{Z}$
denote fermionic and bosonic Matsubara frequencies , respectively.
The electronic and phononic self-energies obey the equations \citep{AGD63},
as diagrammatically shown in Fig.~\ref{Fig2}~(a):

\begin{align}
\Sigma_{{\bf {k}}}(i\omega_{n}) & =-\frac{g_{\text{eff}}^{2}}{\beta N}\sum_{m,{\bf q}}{\cal D}_{{\bf {k}-q}}(i\omega_{n}-i\omega_{m}){\cal G}_{{\bf {q}}}(i\omega_{m}),\label{eq:G}\\
\Pi_{{\bf {k}}}(i\omega_{n}) & =\frac{2g_{\text{eff}}^{2}}{\beta N}\sum_{m,{\bf q}}{\cal G}_{{\bf {q}}}(i\omega_{m}){\cal G}_{{\bf {q}-k}}(i\omega_{m}-i\omega_{n}).\label{eq:D}
\end{align}
These equations define the properties of the normal state of the electron
gas. In order to find the superconducting transition temperature,
we solve the linearized self-consistent equation for the pairing vertex
$\Gamma_{{\bf {k}}}$:

\begin{align}
\Gamma_{{\bf {k}}}\left(i\omega_{n}\right) & =-\frac{g_{\text{eff}}^{2}}{N\beta}\sum_{{\bf {q}}}{\cal D}_{{\bf {k}-q}}(i\omega_{n}-i\omega_{m})\Gamma_{{\bf {q}}}(i\omega_{m})\nonumber \\
 & \times{\cal G}_{{\bf {q}}}(i\omega_{m}){\cal G}_{-{\bf {q}}}(-i\omega_{m}).\label{eq:Phi}
\end{align}
The highest-temperature solution of this equation defines the critical
temperature $T_{c}$. We provide an analytical solution of Eqs. (\ref{eq:G}-\ref{eq:Phi})
under several simplifying assumptions. In particular, we assume that
the detuning $\delta_{{\bf {q}}}$ and the squeezing parameter $D_{{\bf {q}}}$
do not depend on momentum ${\bf {q}}$. In this case, the only momentum
dependence in Eqs.~(\ref{eq:G}, \ref{eq:D}) is due to the electron
polarization operator. The latter Eq.~\eqref{eq:D} contains static
$\omega_{n}=0$ and dynamical contributions $\omega_{n}\neq0$. The
static contribution is responsible for the phonon softening due to
the interaction with electrons and it is generally important at strong
couplings. In addition, it effectively enhances the electron-phonon
interaction \citep{CAE20}. The dynamical contribution describes the
Landau damping. We neglect the dynamical contribution as it is smaller
than the first Matsubara frequency term in the denominator of ${\cal D}$
in the relevant temperature ranges \citep{CAE20}. In addition, we
restrict the polarization operator in Eq.~\eqref{eq:D} to its zeroth
Matsubara component taken with respect to the unperturbed fermionic
Green's function: $\Pi_{{\bf {k}}}\left(i\omega_{n}\right)\approx-2\nu_{0}g_{\text{eff}}^{2}$,
where $\nu_{0}\equiv N^{-1}\sum_{{\bf k}}\delta(E_{\text{F}}-\epsilon_{{\bf k}})$
is the density of states at the Fermi energy $E_{\text{F}}$. We note
that according to this definition, $\nu_{0}$ has the dimension of
inverse energy. Under these assumptions the renormalized phonon propagator
takes the following form:

\begin{equation}
{\cal D}\left(i\omega_{n}\right)=\frac{-2\left(\delta+D\right)}{\omega_{n}^{2}+\left(\delta^{2}-D^{2}\right)\left(1-2\lambda_{0}\right)},\label{eq:D_eff}
\end{equation}
where the effective electron-phonon coupling is defined as $\lambda_{0}=2\nu_{0}g_{\text{eff}}^{2}/\left(\delta-D\right)$.
The ``anti-squeezing'' manifests itself as an excess noise of the
phonon field $\hat{\phi}_{{\bf {q}}}$ which is found by taking the
Matsubara frequency sum in Eq.~\eqref{eq:D_eff} in the $T\rightarrow0$
limit $\langle\hat{\phi}_{{\bf {q}}}\hat{\phi}_{-{\bf {q}}}\rangle=$$-\beta^{-1}\sum_{n}{\cal D}_{{\bf {q}}}(i\omega_{n})\approx\sqrt{(\delta+D)/((1-2\lambda_{0})(\delta-D))}.$
We see that the phonon fluctuations are enhanced by interaction with
electrons and by external squeezing in a multiplicative way.

Since the righthand sides of Eqs.~(\ref{eq:G}, \ref{eq:Phi}) do
not depend on momentum $k$, the dependence can be eliminated by by
taking an average over Fermi surface $\Gamma_{n}\rightarrow\langle\Gamma_{{\bf {k}}}(i\omega_{n})\rangle_{\text{FS}}$,
$\Sigma_{n}\rightarrow\langle\Sigma_{{\bf {k}}}(i\omega_{n})\rangle_{\text{FS}}$.
An approximate analytical solution of these equation is known \cite{AD75,M68},
and yields the following expression for the critical temperature:

\begin{equation}
T_{c}=\frac{\sqrt{\left(\delta^{2}-D^{2}\right)\left(1-2\lambda_{0}\right)}}{1.2}e^{-1.04\frac{\lambda_{\text{eff}}+1}{\lambda_{\text{eff}}}},\label{eq:T_c}
\end{equation}
where the effective coupling strength is defined as $\lambda_{\text{eff}}=\lambda_{0}/\left(1-2\lambda_{0}\right)$
\citep{CAE20}, and the first term in this expression stands for the
effective phonon bandwidth, which corresponds to the poles of Eq.
\eqref{eq:D_eff} with respect to the Matsubara frequency. At strong
coupling, $\lambda_{0}$ the system undergoes a transition to charge-density
phase \cite{CAE20,ENH18,A01}. In Eq~\eqref{eq:T_c}, it manifests
itself as singularity of $\lambda_{\text{eff}}$ at $\lambda_{0}^{\text{cr}}=0.5$.
Due to the vertex corrections neglected in Eqs.~(\ref{eq:G}-\ref{eq:Phi}),
the exact Monte-Carlo treatment of Holstein model \cite{ENH18} predicts
a slightly different value $\lambda_{0}^{\text{cr}}\approx0.4$. In
the following, we will restrict all system parameters such that $\lambda_{0}\leq\lambda_{0}^{\text{cr}}$
in order to avoid this instability.

We now analyze Eq.~\eqref{eq:T_c} expression by varying $\delta$
and $D$ while assuming $g_{\text{eff}}$ is fixed. In the absence
of squeezing ($D=0$), we find the maximum $T_{c}$ with respect to
the detuning $\delta$ being equal to $T_{c}^{\text{max}}\approx0.4g_{\text{eff}}^{2}\nu_{0}$.
This value, being expressed in terms of the optimal detuning, is equal
to $T_{c}^{\text{max}}\approx0.08\delta$, which reproduces the known
result \citep{EKS18,ENH18}. In order to study the influence of squeezing
on the superconducting temperature we assume the squeezing parameter
$D$ is fixed to some positive value. In this case a new maximum with
respect to $\delta$ is straightforwardly found to be $T_{c}^{\text{max}}\left(D\right)\approx0.25\sqrt{g_{\text{eff}}^{2}\nu_{0}D}$
in the limit when $D\gg g_{\text{eff}}^{2}\nu_{0}$. It is achieved
at $\delta^{\text{max}}\approx D+5g_{\text{eff}}^{2}\nu_{0}$. This
combination of squeezing and detuning saturates the the bare electron-phonon
coupling to $\lambda_{0}\approx$$\lambda_{0}^{\text{cr}}$, which
is approximately independent of $D$ and $\delta$. The effective
phonon bandwidth for the optimal detuning scales as $\sqrt{\left(\delta^{2}-D^{2}\right)\left(1-2\lambda_{0}\right)}\propto\sqrt{g_{\text{eff}}^{2}\nu_{0}D}$
which determines the scaling of $T_{c}$ at large squeezing. The enhancement
can therefore be seen as increasing of the effective bandwidth of
phonons while keeping the effective coupling fixed to its maximal
value \citep{ENH18,EKS18}. In general $g_{\text{eff}}$ also depends
on $D$ and $\delta$ due to being proportional to the steady-state
phonon occupation. The presented analysis can be straightforwardly
extended to take this into account.

\begin{figure}
\begin{centering}
\includegraphics[scale=0.6]{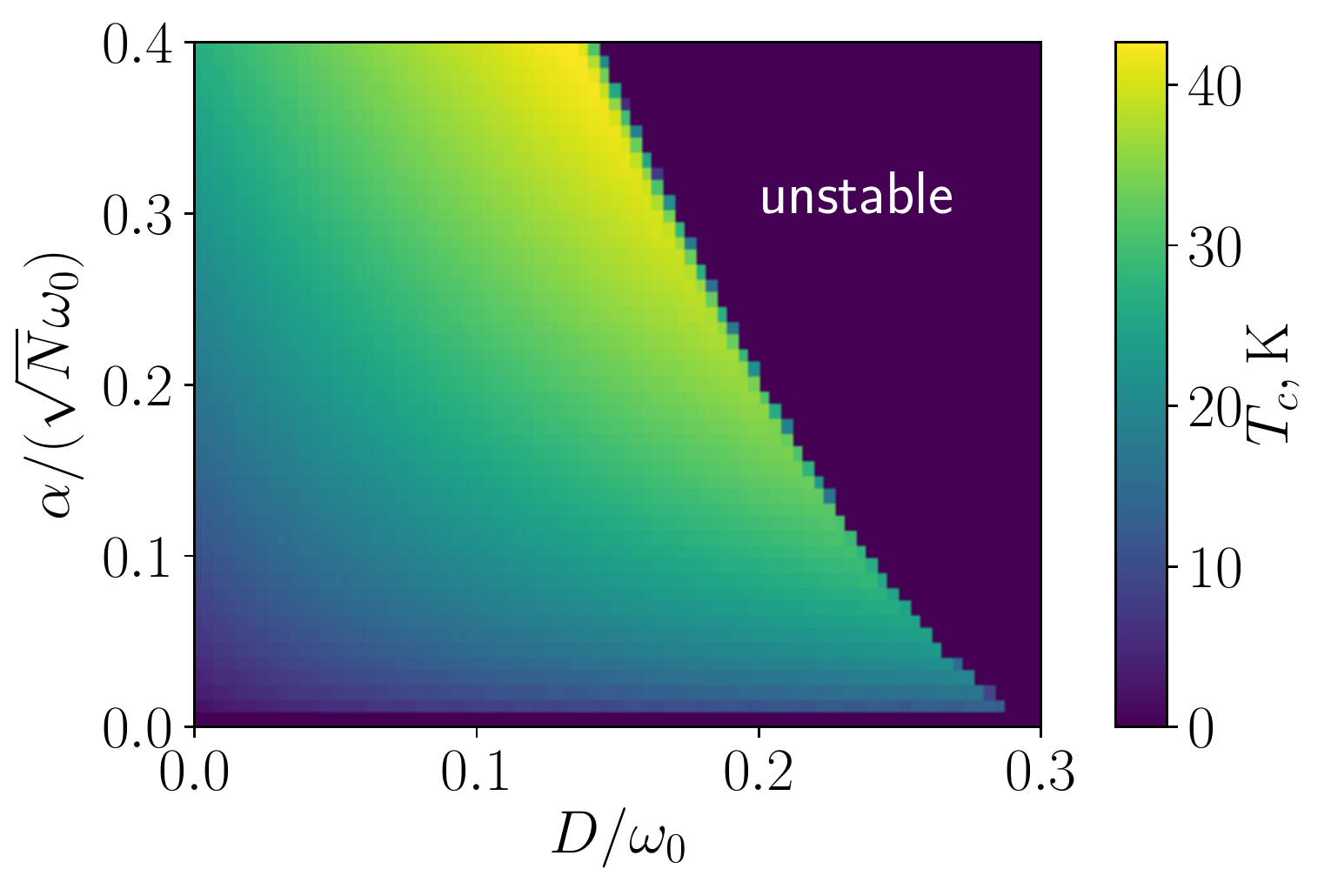} 
\par\end{centering}
\caption{Superconducting critical temperature $T_{c}$ as function of the linear
driving $\alpha$ and parametric driving strength $D$. The detuning
is chosen to produce the strongest effect at each point. Simulation
parameters are the same as in Fig.~\ref{Fig1}. The strongest enhancement
is achieved at the boundary of lattice instability region. }

\label{Fig3} 
\end{figure}

We now compare the analytical prediction for the critical temperature
with the numerical self-consistent solution of Eqs.~(\ref{eq:G},
\ref{eq:Phi}) performed in a discretized $52\times52$ -momentum
and $400$ Matsubara frequency lattice space. We consider the external
driving $\alpha=0.25\omega_{0}$ to be fixed while we vary the detuning
$\delta$. The critical temperature as function of the detuning is
shown in Fig.~\ref{Fig2} for several values of the squeezing parameter
$D$. The smallest detuning of all curves correspond to $\lambda_{0}\approx\lambda_{0}^{\text{cr}}$.
The maximum{ $T_{c}$} is achieved at the lowest possible
$\delta$ in agreement with the analytical expression provided above.
We study the effect of linear $\alpha$ and parametric $D$ driving
on superconducting $T_{c}$ for optimal values of detuning $\delta$
Fig.~\ref{Fig3}. The external squeezing allows one to achieve strong
enhancement at much lower driving intensities $\alpha$, and the strongest
effect is achieved on the boundary of the lattice instability regimes.

We now discuss two possible experimental realizations of our idea
to generate phonon squeezing. The first proposal exploits the intrinsic
photon-phonon coupling nonlinearities \cite{CNF18}. For illustration
purposes, we consider a simplified model of two-dimensional electron
lattice gas with the nearest-neighbor tunneling rate $J\approx0.2\omega_{0}$,
corresponding to the $K_{3}C_{60}$ fulleride superconductor \cite{KWR17,KBR16}.
We assume that the infrared-active phonon mode with Debye frequency
$\omega_{0}\approx0.17\text{eV}$ being driven by a bi-chromatic light
at frequencies $\omega_{p}$ and $2\omega_{p}$. For an estimate of
achievable phonon parametric driving rate, we take the values achieved
with phonon parametric amplification \cite{CNF18} $D\propto0.1\omega_{0}$.
Here we focus only on pairing induced by external driving. In our
simulations we consider the electron-phonon coupling coefficient $g\approx0.1\omega_{0}$
\cite{KWR17} and the electron density of states $\nu_{0}\approx0.6\omega_{0}^{-1}$.
We note that in deriving Hamiltonian (\ref{eq:H_ph}, \ref{eq:H_int})
we neglected the terms rotating at $\propto2\omega_{0}$ \cite{GSB20}
(see SM) which may induce heating for broader-band materials. Our
simplified analysis can be extended with taking these rotating terms
into account perturbatively \cite{KBR16}. With the parameters above
we estimate the bare electron-phonon coupling $g^{2}\nu_{0}\approx0.006\omega_{0}$.
The corresponding values of the critical temperature are shown in
Fig.~\ref{Fig1}~(b).

In the second approach, the phonon squeezing can be injected by coupling
a 2-dimensional \citep{LWB20} superconductor optomechanical membrane
coupled to a parametrically-driven cavity mode Fig.~\ref{Fig1} (a).
The main challenge in this case is to control the high-frequency phonons
as the critical temperature Eq.~\eqref{eq:T_c} is proportional to
the overall frequency range of the phonon modes. Coupling of light
to high-frequency phonons has been demonstrated in several setups
including the optomechanical disk resonators \citep{DBS11}, and high-frequency
bulk acoustic phonons\citep{EKB19}. The squeezing of phonons can
be achieved via hybridization with photons which are parametrically
driven \citep{GLL20,GH16}. By assuming frequency range of the order
of 100 GHz, as achieved in resonators based on acoustic distributed
Bragg reflectors \citep{CBT14}, we can estimate the enhancement to
be of the order of $T_{c}\propto3K$ for the same parameter ratio
as provided in the previous paragraph.

The main limiting factor is the effective phonon bandwidth which is
substantially reduced due to squeezing close to the parametric instability.
However, the analysis presented in this paper is restricted to the
the isotropic case i.e. when $D_{{\bf {q}}}=D,$ $\delta_{{\bf {q}}}=\delta$.
The momentum dependance of the $D_{{\bf {q}}}$ and $\delta_{{\bf {q}}}$,
which is generally present in experiment, provide an additional degree
of freedom. In particular, control of the effective phonon dispersion
independently of the coupling strength.

In conclusion, we studied the enhancement of superconductivity due
to an externally induced squeezing. The phase-sensitive squeezing
enhances quadrature fluctuations of the phonon field leading to exponentially
stronger interaction, while reducing the spectral bandwidth of phonons.
We study the competition of these two effects numerically and analytically
and find a parameter range of enhanced superconductivity. The effective
squeezed Holstein model describing the system allows also to dynamically
suppress coupling to a certain range of phonon modes. The strength
of the suppression is exponential. This can be very useful in case
when superconductivity competes with other types of instabilities
e.g. charge-density wave instability. By decoupling from the phonon
modes responsible for the instability, one can enhance the superconducting
transition. This opens up a way to engineer an effective electron-phonon
interacting model which suppress polaronic/CDW tendencies.

\paragraph*{Acknowledgments}

{The authors acknowledge useful discussions with Ivar Martin and
Dmitry Basov. This work was supported by ARO W911NF2010246 (A.G.),
ARO W911NF2010232 (M.H.), NSF DMR-1613029, ARO W911NF1310172, and
Simons Foundation (M.H., V.G.).}

\appendix
 \bibliographystyle{apsrev4-1}
\bibliography{bibl.bib}

%
%
%
%
%
%
%
%
%
%
%
%
%
%
%
%
%
\clearpage
\onecolumngrid
\appendix*

\subsection{Enhancement in ensemble of two-level systems}

To illustrate the effect of squeezing on the mediated interaction and the phase-sensitive nature of the increase/decrease of the interaction strength, we study the case of spins coupled to a bosonic modes. While the focus of the main text is on electrons in 2D, here we present the key idea using a toy model with spins.  Moreover, such a phase-sensitive nature was not discussed in previous works on the subject \cite{zeytinouglu2017engineering}.  To be concrete,    we study interaction of an ensemble of $N$ two-level
systems (TLS) with the squeezed bosonic mode. We demonstrate a phase-sensitive
enhancement analogous to those studied in the main text. We first
define the raising/lowering operators for TLS as $\sigma_{i}^{\pm}$
and the bosonic annihilation operator as $a$. We consider the following
Hamiltonian:

\begin{align*}
H & =\omega_{0}a^{\dagger}a+\omega_{a}\sum_{i=1}^{N}\sigma_{i}^{+}\sigma_{i}^{-}\\
 & +\frac{D}{2}\left(e^{2i\phi_{D}}a^{2}e^{2i\omega_{a}t}+e^{-2i\phi_{D}}e^{-2i\omega_{a}t}a^{\dagger2}\right)\\
 & +g\left(a+a^{\dagger}\right)\sum_{i=1}^{N}\left(\sigma_{i}^{+}+\sigma_{i}^{-}\right)
\end{align*}
where $\omega_{a}$ is the transition energy of a two-level system,
$\omega_{0}$ is the frequency of a bosonic mode and $D$ denotes
the squeezing rate. We transform into rotating frame $a\rightarrow ae^{-i\omega_{a}t-i\phi_{D}}$
and $\sigma_{i}^{-}=\sigma_{i}^{-}e^{-i\omega_{a}t}$. By denoting
the detuning $\delta=\omega_{0}-\omega_{a}$ we get and performing
the rotating-wave approximation (RWA):

\begin{align*}
H & =\delta a^{\dagger}a+\frac{D}{2}\left(a^{2}+a^{\dagger2}\right)+g\left(e^{-i\phi_{D}}aJ^{+}+e^{i\phi_{D}}a^{\dagger}J^{-}\right),
\end{align*}
where we denoted $J^{\pm}=\sum\sigma_{i}^{\pm}$, $\left[J^{+},J^{-}\right]=2J^{z}$.
By performing the Bogolyubov transformation $a=ub+vb^{\dagger}$ with
$u=\cosh\left[r\right],v=\sinh\left[r\right]$, $r=\text{arctanh}\left[-D/\delta\right]$
we get:

\begin{align*}
H & =\sqrt{\delta^{2}-D^{2}}a^{\dagger}a\\
 & +g(b\{e^{-i\phi_{D}}u\sum_{i}\sigma_{i}^{+}+e^{i\phi_{D}}v\sum_{i}\sigma_{i}^{-}\}+\text{H.c.})
\end{align*}
We now perform adiabatic elimination of bosonic mode by means of the
Schrieffer-Wolff transformation:

\begin{equation}
H^{\prime}=e^{-S}He^{S}=H+\frac{1}{2}\left[H,S\right]+\ldots\label{eq:Hprime}
\end{equation}
with

\begin{align*}
S & =\frac{2g}{\sqrt{\delta^{2}-D^{2}}}\left(b\left\{ e^{-i\phi_{D}}uJ^{+}+e^{i\phi_{D}}vJ^{-}\right\} \right.\\
 & -\left.b^{\dagger}\left\{ e^{-i\phi_{D}}vJ^{+}+e^{i\phi_{D}}uJ^{-}\right\} \right),
\end{align*}
By taking the necessary commutators in \eqref{eq:Hprime} we get:

\begin{align*}
H^{\prime} & =-g^{2}\left\{ \frac{e^{-2i\phi_{D}}D}{\delta^{2}-D^{2}}J^{+2}+\frac{De^{2i\phi_{D}}}{\delta^{2}-D^{2}}J^{-2}\right\} \\
 & -g^{2}\frac{\delta}{\delta^{2}-D^{2}}\left(J^{+}J^{-}+J^{-}J^{+}\right)\\
 & -\frac{g^{2}}{\sqrt{\delta^{2}-D^{2}}}\left(2b^{\dagger}b+1\right)J^{z}
\end{align*}
In order to represent this result in a more physically-appealing form,
we transform variables as $J^{\pm}e^{\mp i\phi_{D}}\rightarrow J^{\pm}$
and denoting $J^{x/y}=\frac{1}{2}\left(J^{+}\pm iJ^{-}\right)$ we
find:

\begin{align*}
H^{\prime} & =-g^{2}\left\{ \frac{1}{\delta-D}\left(J^{x}\right)^{2}+\frac{1}{\delta+D}\left(J^{y}\right)^{2}\right\} \\
 & -\frac{g^{2}}{\sqrt{\delta^{2}-D^{2}}}\left(2b^{\dagger}b+1\right)J^{z}
\end{align*}
We therefore find that interaction is enhanced in one quadrature and
decreased in the other. We note that the definition of $J^{x/y}$
is arbitrary without external reference.

\subsection{Derivation of the effective Hamiltonian\label{sec:Derivation-of-the}}

In this section, we provide technical details of the derivation of
the effective Hamiltonian Eqs.~(4, 5) of the main text. We treat
phonon dynamics in terms of mean-field theory and keep quadratic fluctuations.
We represent phonon operators as $\hat{\phi}_{{\bf {q}}}=\langle\hat{\phi}_{0}\rangle\delta_{{\bf {q},0}}+\tilde{\hat{\phi}}_{{\bf {q}}}$.
The mean-field set of equations reads:

\begin{align}
\frac{d}{dt}\langle\hat{\phi}_{0}\rangle & =\omega_{0}\langle\hat{\pi}_{0}\rangle,\label{eq:MF}\\
\frac{d}{dt}\langle\hat{\pi}_{0}\rangle & =-\omega_{0}\langle\hat{\phi}_{0}\rangle-4\alpha\cos\left(\omega_{p}t+\psi_{\alpha}\right)\nonumber \\
 & -4\langle\hat{\phi}_{0}\rangle D_{0}\cos\left(2\omega_{p}t+\psi_{D}\right),\label{eq:MF2}
\end{align}
where we defined the ``momentum'' operator $\hat{\pi}_{0}\equiv i(a_{0}^{\dagger}-a_{0}$).
In deriving these equations we neglected coupling to the electrons.
The latter may induce frequency shift. The resulting effective Hamiltonian
for fluctuating part is (we omit tildes for shortness):

\begin{align}
H_{\text{ph}} & =\sum_{{\bf {q}}}\omega_{{\bf {q}}}\hat{a}_{{\bf {q}}}^{\dagger}\hat{a}_{{\bf q}}+\sum_{{\bf {q}}}D_{{\bf {q}}}\cos\left(2\omega_{p}t+\psi_{D}\right)\hat{\phi}_{{\bf {q}}}\hat{\phi}_{-{\bf {q}}},\\
H_{\text{int}} & =\frac{2g}{N}\sum_{\sigma,k,q}\hat{c}_{{\bf {k}+q,\sigma}}^{\dagger}\hat{c}_{{\bf {k},\sigma}}\hat{\phi}_{{\bf {q}}}\langle\hat{\phi}_{0}\rangle\nonumber \\
 & +\frac{g}{N}\sum_{\sigma,k,q}\hat{c}_{{\bf {k},\sigma}}^{\dagger}\hat{c}_{{\bf {k},\sigma}}\langle\hat{\phi}_{0}\rangle^{2},\label{eq:H_intSM}
\end{align}
where we neglected the quadratic in $\hat{\phi}$ electron-phonon
coupling in $H_{\text{int}}$. The second term in Eq.~\eqref{eq:H_intSM}
stands for the renormalization of the chemical potential. As shown
in section below, the mean field value $\langle\hat{\phi}_{0}\left(t\right)\rangle\approx2\text{Re}[\bar{a}_{0}e^{-i\omega_{p}t}]$,
where 
\[
\bar{a}_{0}=\alpha\frac{D_{0}e^{i\left(\psi_{\alpha}-\psi_{D}\right)}-\delta_{0}e^{-i\psi_{\alpha}}}{\left(\delta_{0}^{2}-D_{0}^{2}\right)}.
\]
We now transform into the frame rotating at $\omega_{p}$ and neglect
all high-frequency rotating terms:

\begin{align}
H_{\text{ph}} & =\sum_{{\bf {q}}}\left(\omega_{{\bf {q}}}-\omega_{p}\right)a_{{\bf {q}}}^{\dagger}\hat{a}_{{\bf q}}\nonumber \\
 & +\sum_{{\bf {q}}}\frac{D_{{\bf {q}}}}{2}\left(e^{i\psi_{D}}\hat{a}_{{\bf q}}a_{-{\bf {q}}}+e^{-i\psi_{D}}a_{-{\bf {q}}}^{\dagger}a_{{\bf {q}}}^{\dagger}\right)\label{eq:H1-1-1}\\
H_{\text{int}} & =\frac{2g}{N}\sum_{\sigma,k,q}c_{{\bf {k}+q,\sigma}}^{\dagger}c_{{\bf {k},\sigma}}\left(\hat{a}_{{\bf q}}\bar{a}_{0}^{*}+a_{-{\bf {q}}}^{\dagger}\bar{a}_{0}\right).\label{eq:H3-1-1}
\end{align}
We recover find an effective Holstein model provided in the main text.
In deriving the Hamiltonian \eqref{eq:H1-1-1},\eqref{eq:H3-1-1}
we neglected the following terms rotating at $2\omega_{p}$:

\begin{align*}
H^{\text{rot}} & =\sum_{{\bf {q}}}\frac{D_{{\bf {q}}}}{2}\left(e^{i\left(2\omega_{p}t+\psi_{D}\right)}+e^{-i\left(2\omega_{p}t+\psi_{D}\right)}\right)\left(a_{{\bf q}}a_{{\bf q}}^{\dagger}+a_{-{\bf q}}^{\dagger}a_{-{\bf q}}\right),\\
 & +\frac{2g}{N}\sum_{\sigma,k,q}\hat{c}_{{\bf {k}+q,\sigma}}^{\dagger}\hat{c}_{{\bf {k},\sigma}}\left(a_{{\bf q}}\bar{a}_{0}e^{-2i\omega_{p}t}+a_{-{\bf q}}^{\dagger}\bar{a}_{0}^{*}e^{2i\omega_{p}t}\right)\\
 & +\frac{g}{N}\sum_{\sigma,k,q}\hat{c}_{{\bf {k},\sigma}}^{\dagger}\hat{c}_{{\bf {k},\sigma}}\left(\bar{a}_{0}^{2}e^{-2i\omega_{p}t}+\bar{a}_{0}^{*2}e^{2i\omega_{p}t}\right).
\end{align*}
This approximation is valid as soon as there are no possible resonant
transitions in the electron gas caused by the rotating terms. 

\subsection{Steady state phonon properties\label{sec:Steady-state-phonon}}

Here we consider properties steady state properties of phonon modes.
We first approximately solve the mean field set of equations Eqs.~(\ref{eq:MF},
\ref{eq:MF2}). For that we assume $\langle\hat{\phi}_{0}\left(t\right)\rangle\approx2\text{Re}\left[\bar{a}_{0}e^{-i\omega_{p}t}\right]$
and find $\bar{a}_{0}$ neglecting coupling to higher frequency components.
We note that by making this ansatz we neglect terms rotating at $3\omega_{p},5\omega_{p},\ldots$
which produce only rapidly rotating terms in Eqs.~(\ref{eq:MF},
\ref{eq:MF2}).

This results in the following equations:

\begin{align*}
\left(\omega_{0}^{2}-\omega_{p}^{2}\right)\bar{a}_{0} & =-2D_{0}\omega_{0}\bar{a}_{0}^{*}e^{-i\theta_{D}}-2\alpha\omega_{0}e^{-i\theta_{\alpha}}\\
\left(\omega_{0}^{2}-\omega_{p}^{2}\right)\bar{a}_{0}^{*} & =-2D_{0}\omega_{0}\bar{a}_{0}e^{i\theta_{D}}-2\alpha\omega_{0}e^{i\theta_{\alpha}}
\end{align*}

The solution is: 
\begin{align*}
\bar{a}_{0} & =\alpha\frac{4D_{0}\omega_{0}^{2}e^{i\left(\theta_{\alpha}-\theta_{D}\right)}-2\omega_{0}\left(\omega_{0}^{2}-\omega_{p}^{2}\right)e^{-i\theta_{\alpha}}}{\left(\left(\omega_{0}^{2}-\omega_{p}^{2}\right)^{2}-4D_{0}^{2}\omega_{0}^{2}\right)}
\end{align*}

Using $\omega_{0}=\omega_{p}+\delta_{0}$ and expanding in the limit
$\omega_{p}/\delta_{0}\rightarrow\infty$ we find: 
\[
\bar{a}_{0}\approx\alpha\frac{D_{0}e^{i\left(\theta_{\alpha}-\theta_{D}\right)}-\delta_{0}e^{-i\theta_{\alpha}}}{\left(\delta_{0}^{2}-D_{0}^{2}\right)}
\]

\paragraph{Phonon propagator}

In Eq.~\eqref{eq:H3-1-1} electrons are effectively coupled to the
phonon field $\hat{\Phi}_{{\bf {q}}}=\left(\hat{a}_{{\bf q}}e^{-i\arg(a_{0})}+a_{-{\bf {q}}}^{\dagger}e^{i\arg(a_{0})}\right)$
. We now derive bare propagator of this field

\[
{\cal D}^{R}\left[\omega\right]=-i\int_{0}^{\infty}e^{i\omega t}\langle[\Phi_{{\bf {q}}}\left(t\right),\Phi_{-{\bf {q}}}\left(0\right)]\rangle
\]
We start with the set of Heisenberg equations of motion with respect
to the bare phonon Hamiltonian Eq.~\eqref{eq:H1-1-1}:

\begin{align*}
\frac{d}{dt}\hat{a}_{{\bf q}} & =-i\delta_{{\bf q}}\hat{a}_{{\bf q}}-iD_{{\bf {q}}}e^{-i\theta_{D}}a_{-{\bf {q}}}^{\dagger}\\
\frac{d}{dt}a_{-{\bf {q}}}^{\dagger} & =i\delta_{{\bf q}}a_{-{\bf {q}}}+iD_{{\bf {q}}}e^{i\theta_{D}}\hat{a}_{{\bf q}}
\end{align*}
Solving them we find the imaginary-time propagator:

\[
{\cal D}_{{\bf {q}}}\left[i\omega_{n}\right]=\frac{2(D_{{\bf {q}}}\cos(\theta_{D}+2\arg(a_{0}))-\delta_{{\bf q}})}{\omega_{n}^{2}+\delta^{2}-D_{{\bf {q}}}^{2}}
\]
Numerator of this expression is maximized by e.g. the following choice
of driving phases $\phi_{D}=\pi$ and $\psi_{\alpha}=0$:

\[
{\cal D}_{{\bf {q}}}\left[i\omega_{n}\right]=\frac{-2(D_{{\bf {q}}}+\delta_{{\bf q}})}{\omega_{n}^{2}+\delta_{{\bf q}}^{2}-D_{{\bf {q}}}^{2}}
\]

\subsection{Eliashberg equations\label{sec:Eliashberg-equations}}

By denoting $\Sigma\left[k,i\omega_{n}\right]\equiv i\omega_{n}\left[1-Z\left[k,i\omega_{n}\right]\right]$
in the particle-hole symmetric case and averaging Eqs.~(6-8) over
the Fermi surface $\Gamma_{n}\rightarrow\left\langle \Gamma_{{\bf {k}}}\left(\omega_{n}\right)\right\rangle _{\text{FS}}$,
$Z_{n}\rightarrow\left\langle Z_{{\bf {k}}}\left(\omega_{n}\right)\right\rangle _{\text{FS}}$
we get \citep{M20,CAE20}:

\begin{align*}
\Gamma_{n} & =\frac{\pi}{\beta}\sum_{m}\lambda_{\text{eff}}\left[i\omega_{n}-i\omega_{m}\right]\frac{\Gamma_{m}}{Z_{m}\left|\omega_{m}\right|},\\
Z_{n} & =1+\frac{\pi}{\beta}\sum_{m}\lambda_{\text{eff}}\left[i\omega_{n}-i\omega_{m}\right]\left(\frac{\omega_{m}}{\left|\omega_{m}\right|}\right),
\end{align*}
where $\lambda_{\text{eff}}\left[i\omega_{n}-i\omega_{m}\right]=\nu_{0}g_{\text{eff}}^{2}{\cal D}\left[\vartheta,i\omega_{n}-i\omega_{m}\right]$.

\subsection{Multimode cavity optomechanics}

In this section we show that the effective parametric driving of the
out-of-plane phonons Eq. (4) in a multimode cavity optomechanical
setting. Mathematical formalism is essentially an extension of \citep{GH16,GLL20,SLY20}
to a multimode cavity case. By denoting the annihilation operator
of photon field with the transverse momentum $\text{k}$ as ${\cal E}_{{\bf k}}$,
the photon-phonon Hamiltonian reads:

\begin{align}
H_{\text{int}} & =\frac{g_{\text{ph}}}{\sqrt{N}}\sum_{{\bf k},{\bf q}}{\cal E}_{{\bf k+q}}^{\dagger}{\cal E}_{{\bf k}}\phi_{{\bf q}}\\
H_{\text{phon}} & =\sum_{{\bf q}}\omega_{{\bf q}}a_{{\bf q}}^{\dagger}a_{{\bf q}}\\
H_{\text{phot}} & =\omega_{\text{cav}}\sum_{{\bf k}}{\cal E}_{{\bf k}}^{\dagger}{\cal E}_{{\bf k}}+\zeta\left({\cal E}_{0}e^{i\Omega_{1}t}+{\cal E}_{0}^{\dagger}e^{-i\Omega_{1}t}\right)
\end{align}
where $g_{\text{ph}}$ is the photon-phonon coupling constant, $\zeta$
stands for the cavity driving rate, $\omega_{\text{cav}}$ is the
cavity frequency. In addition, we assume that photons are parametrically
driven at the frequency $2\Omega_{2}$ with the driving strength $\xi_{{\bf k}}$:

\begin{equation}
H_{\text{par}}=\sum_{{\bf k}}\frac{\xi_{{\bf k}}}{2}\left({\cal E}_{{\bf k}}{\cal E}_{{\bf -k}}e^{2i\Omega_{2}t}+{\cal E}_{{\bf k}}^{\dagger}{\cal E}_{{\bf -k}}^{\dagger}e^{-2i\Omega_{2}t}\right)
\end{equation}
We now linearize the interaction Hamiltonian assuming the driving
is strong enough and get:

\begin{equation}
H_{\text{int}}=\frac{\zeta_{\text{ss}}g_{\text{ph}}}{\sqrt{N}}\sum_{{\bf q}}\left({\cal E}_{{\bf q}}^{\dagger}e^{-i\Omega_{1}t}+{\cal E}_{-{\bf q}}e^{i\Omega_{1}t}\right)\phi_{{\bf q}}
\end{equation}
where the mean-field cavity coherence is $\left\langle {\cal E}_{0}\right\rangle \approx\zeta_{\text{ss}}e^{-i\Omega_{1}t}$
with $\zeta_{\text{ss}}=-\zeta/(\omega_{\text{cav}}-\Omega_{1})$.
We now transform to the interaction picture: ${\cal E}_{{\bf q}}\rightarrow{\cal E}_{{\bf q}}e^{-i\Omega_{2}t}$
and $a_{{\bf q}}\rightarrow a_{{\bf q}}e^{i\left(\Omega_{2}-\Omega_{1}\right)t}$
and neglect all rotating terms:

\begin{align}
H_{\text{int}} & \approx\frac{\zeta_{\text{ss}}g_{\text{ph}}}{\sqrt{N}}\sum_{{\bf q}}\left({\cal E}_{{\bf q}}^{\dagger}a_{{\bf q}}+{\cal E}_{{\bf q}}a_{{\bf q}}^{\dagger}\right)\\
H_{\text{phon}} & =\sum_{{\bf q}}\delta_{q}a_{{\bf q}}^{\dagger}a_{{\bf q}}\\
H_{\text{phot}} & =\Delta_{\text{cav}}\sum_{{\bf k}}{\cal E}_{{\bf k}}^{\dagger}{\cal E}_{{\bf k}}\\
H_{\text{par}} & =\sum_{{\bf k}}\frac{\xi_{{\bf k}}}{2}\left({\cal E}_{{\bf k}}{\cal E}_{{\bf -k}}+{\cal E}_{{\bf k}}^{\dagger}{\cal E}_{{\bf -k}}^{\dagger}\right)
\end{align}
where $\Delta_{\text{cav}}=\omega_{\text{cav}}-\Omega_{2}$ and $\delta_{q}=\left(\omega_{{\bf q}}-\Omega_{1}+\Omega_{2}\right)$.
Adiabatic elimination of cavity modes. We First do the Bogolyubov
transform: ${\cal E}_{{\bf k}}=u_{q}{\cal \gamma}_{{\bf q}}^{\dagger}+v_{q}\gamma_{-q}$
and $\Lambda_{{\bf k}}=\sqrt{\Delta_{\text{cav}}^{2}-\xi_{{\bf k}}^{2}}$.
And second, we eliminate $\gamma_{{\bf q}}$ assuming it is in vacuum
state. For $\Lambda_{{\bf k}}\gg\delta_{{\bf q}}:$

\begin{align*}
H_{\text{phon}} & =\sum_{{\bf q}}\left(\delta_{q}+\frac{\Delta_{\text{cav}}}{\Delta_{\text{cav}}^{2}-\xi_{{\bf q}}^{2}}\frac{\zeta_{\text{ss}}^{2}g_{\text{ph}}^{2}}{N}\right)a_{{\bf q}}^{\dagger}a_{{\bf q}}\\
 & +\frac{\zeta_{\text{ss}}^{2}g_{\text{ph}}^{2}}{2N}\sum_{{\bf q}}\frac{\xi_{{\bf q}}}{\Delta_{\text{cav}}^{2}-\xi_{{\bf q}}^{2}}\left(a_{-{\bf q}}a_{{\bf q}}+a_{{\bf q}}^{\dagger}a_{{\bf -q}}^{\dagger}\right)
\end{align*}

%
%

\end{document}